\documentclass{article}[12pt]
\usepackage{epsf}
\usepackage[latin1]{inputenc}
\usepackage{amsmath,amsfonts,amssymb}

\def\beq{\begin{equation}}
\def\eeq{\end{equation}}
\def\bea{\begin{eqnarray}}
\def\eea{\end{eqnarray}}
\def\nn{\nonumber}

\def\neq{\not=}

\date{}

\begin{document}
\begin{titlepage}
\begin{center}
{\large\bf Critical interfaces of the Ashkin-Teller model at the parafermionic point.}\\[.3in] 

{\bf M.\ Picco$^{1}$ and R.\ Santachiara$^{2}$}\\
$^1$ {\it LPTHE\/}\footnote[3]{Unit\'e mixte de recherche du CNRS 
UMR 7589.}, 
        {\it  Universit\'e Pierre et Marie Curie-Paris6\\
              Bo\^{\i}te 126, Tour 24-25, 5 \`eme \'etage, \\
              4 place Jussieu,
              F-75252 Paris CEDEX 05, France, \\
    e-mail: {\tt picco@lpthe.jussieu.fr}. }\\
  $^2$ {\it LPTMS\footnote[4]{Unit\'e mixte de 
             recherche du CNRS UMR 8626}, 
             Universit\'e Paris-Sud\\
             B\^atiment 100\\
          F-91405 Orsay, France, \\
    e-mail: {\tt raoul.santachiara@lptms.u-psud.fr}. }\\
\end{center}
\centerline{(Dated: \today)}
\vskip .2in
\centerline{\bf ABSTRACT}
\begin{quotation}

  We present an extensive study of interfaces defined in the $Z_4$
  spin lattice representation of the Ashkin-Teller (AT)
  model. In particular, we numerically compute the fractal dimensions
  of boundary and bulk interfaces at the Fateev-Zamolodchikov
  point. This point is a special point on the self-dual critical line
  of the AT model and it is described in the continuum limit by the ${\cal
    Z}_4$ parafermionic theory. Extending on previous analytical and
  numerical studies \cite{PS,R}, we point out the existence of three
  different values of fractal dimensions which characterize different
  kind of interfaces. We argue that this result may be related to the
  classification of primary operators of the parafermionic algebra.
  The scenario emerging from the studies presented here is expected to
  unveil general aspects of geometrical objects of critical AT model,
  and thus of $c=1$ critical theories in general.

\vskip 0.5cm 
\noindent
\end{quotation}
\end{titlepage}
\section{Introduction and motivation}

The study of the scaling limit of interfaces in systems at critically
has been show to be a very fruitful field of investigation which
provided deep insights in the comprehension of critical phenomena 
\cite{Duplantier}. The richness of conformal symmetry in two dimension
(2D) make the two-dimensional critical systems an ideal framework to
study these issues.  In particular, there is a variety of 2D critical
models for which exact methods of conformal field theory (CFT),
combined with an available Coulomb-gas representation
\cite{Nienhuis_CG}, allow the exact computation of all geometrical
exponents characterizing the fractal shape of critical interfaces
\cite{SalDupl}. Among these models, we mention for instance the
critical percolation, the self-avoiding walks, the loop erased random
walks, or again spin lattice models such as the Potts models \cite{Duplantier}.
 These studies have benefited from a great amount of
numerical work \cite{JS} supporting the proposed theoretical
scenario.  In general, the critical models whose geometrical
properties are well understood, even if there are often no rigorous
proofs, can be associated to the critical phases of a one-parameter
family of statistical models, the $O(n)$ loop models, the parameter
$n$ representing the loop fugacity.  A remarkable recent development
came with the introduction of the Schramm-Loewner evolution (SLE)
which constitutes a family of conformally invariant stochastic growth
process characterized by a parameter $\kappa$ \cite{Review}. The SLE approach offers
a conceptually new description of certain boundary interfaces defined
in the $O(n)$ models.

The critical points of 2D systems can be classified according to
different CFT families.  Each family is characterized by a given set
of (infinite) symmetries and the associated chiral current algebra.
The representation theory of these algebras is at the basis of CFT
constructions. The critical phases of the $O(n)$ models are described
by the most simple of such CFT family, i.e. the one associated to the
conformal symmetry alone and thus to the corresponding Virasoro
algebra \cite{DiFrancesco}.  The connection between this family of
CFTs and SLE approach has been fully understood \cite{BB}.  However,
there are other families of CFTs, the so called extended CFTs, which,
besides the conformal symmetry, enjoy additional infinite symmetries.
These theories describe universality classes which are different from
the ones of the $O(n)$ models. A variety of statistical lattice models
described by extended CFT have been introduced and studied since long
time.  Typically, these models are characterized by symmetries of some
internal degrees of freedom, such as for instance the $SU(2)$ spin
rotational symmetry in the quantum $1+1$ spin chains
\cite{Affleck}. The role of the additional symmetries, in particular
for the so called rational CFT (RCFT), is well understood for many
aspects, as, for instance, the operator algebra of the primary
operators, or the classification of the conformal boundary
conditions. Nevertheless, despite all the recent activity and
progress, the geometrical properties of critical interfaces defined in
such extended CFT are in general not understood.  Moreover, an SLE
approach to describe extended CFT is not known.  For this respect, the
$Z_N$ spin models offer an ideal laboratory to study these
issues. These are lattice model of spins which take $N$ values and
interact via a nearest-neighbor potential which is invariant under a
$Z_N$ cyclic permutation of the $N$ spin states.  The $Z_N$ spin
models admit critical points, the so called Fateev-Zamolochikov (FZ)
points, described by the parafermionic CFT which are extended RCFTs
with $Z_N$ symmetry \cite{FZ}.  The cases with $N = 2$ and $N = 3$ correspond
respectively to the Ising and the three-states Potts model, which in
turn are related to the critical phases of the $O(\sqrt{2})$ and
$O(\sqrt{3})$ loop models.  This is also manifest in the fact that the
associated ${\cal Z}_2$ and ${\cal Z}_3$ parafermionic theories
coincide with the $c=1/2$ and $c=4/5$ Virasoro minimal model $M_3$ and
$M_5$ \cite{DiFrancesco}.  The role of the $Z_N$ symmetry becomes
instead crucial for $N\geq 4$ where the Virasoro algebra is not rich
enough to describe the corresponding parafermionic theory.

 In the $Z_N$ spin lattice models, the boundary and bulk interfaces
 can be naturally defined and the role of the $Z_N$ internal degree of
 freedom is quite explicit in their definition.  Moreover the $Z_N$
 spin models are simple models to be studied numerically.  For $N\geq
 4$, a description of the spin interfaces in terms of a low-energy
 effective field theory is not known, as it is the case for the
 interfaces defined in the $O(n)$ models. For this reason, the
 geometric description of the $Z_N$ spin models are for many aspects
 unknown.  The study of the $Z_N$ spin models at the FZ point is thus
 expected to provide general deep insights on the geometrical
 description of extended CFTs.
 
Numerical measurements of the fractal dimensions associated to the
spin interfaces for $Z_4$ and $Z_5$ spin models at the FZ point were
presented in \cite{PS,PSS}.  In \cite{PS} we considered these $Z_4$
and $Z_5$ spin models on a bounded domain and we investigated the
properties of a boundary interface related to certain boundary
conditions. The numerical results were in the agreement with the
theoretical predictions in \cite{R} where the fractal dimension of
this interface on the basis of the hypothesis of an SLE with an
additional stochastic motion in the internal group of symmetry.  This
approach has been inspired by previous work on the connection between
SLE and CFT with superconformal symmetries \cite{Rasmussen3} or with
additional Lie-group symmetries, \cite{Rasmussen3, Ludwig}.
 
In order to further investigate the geometrical properties of $Z_N$
spin models and the possible consistency with some proposed theoretical
scenario, we studied systematically in \cite{PSS} the bulk geometrical
properties of spin and random cluster interfaces for the $Z_4$ and
$Z_5$ models. These results clearly marked a difference in the
behavior of these non local objects compared to the Ising or the
three-states Potts model.  Among these results there was the observation that the
fractal dimension of certain spin bulk interfaces were different from
the ones corresponding for the boundary interfaces in \cite{PS}.

The $Z_4$ spin model is particularly interesting as it coincides with
the Ashkin-Teller (AT) model.  In \cite{CLR} the fractal dimension of
a boundary interface was investigated along the critical line of the
AT.  At the FZ point of the $Z_4$ spin model, the value of the fractal
dimension was found consistent with the value of the interface studied
in \cite{R} and \cite{PS}.

In this paper we consider in great detail the spin cluster interfaces of the
$Z_4$ spin model at the FZ point by studying systematically different
bulk and boundary interfaces. In order to interpret the numerical
results, in particular in the light of their universal character, we
discuss the classification of the ${\cal Z}_4$ conformal boundary
conditions in terms of boundary spin configurations.

The main result of this paper is the computation of three different
values of fractal dimensions which may encode universal geometrical
properties of the $Z_4$ FZ point.  Even if this results lacks of a
clear theoretical explanation, we point out, on the basis of the
properties of the ${\cal Z}_4$ CFT, a possible scenario for the
geometrical properties of this extended CFT.  It is important to
stress that the ${\cal Z}_4$ CFT coincides with a free Gaussian field
compactified on an orbifold. The study of the AT interfaces complement
thus the results known for the free Gaussian field compactified on a
circle \cite{schramm,Carduslekp,christian} and may suggest an emergent
general behavior for critical interfaces in $c=1$ critical theories.

\section{The model}

We first define  the $Z_4$ spin model. On  each site of 
a square lattice  there is a spin which can take 4 values, $S_i=1,\cdots, 4$. 
The  Hamiltonian defining the model under consideration can be written as 
\begin{equation}
H_{Z_4} =-\sum_{<ij>} A  \delta_{ S_i ,S_j } + B \delta_{S_i ,S_j \pm 1}  + C \delta_{S_i ,S_j \pm 2} \; ,
\label{Hspin}
\end{equation}
where $A,B,C$ are real nonnegative coefficients, $\delta_{S_i,S_j} =
1$ if $S_i=S_j \mod 4$ and $0$ otherwise. Besides a constant
irrelevant term, the interaction is described by two independent real
parameters. The cyclic $Z_4$ symmetry of the model (\ref{Hspin}) is
completely manifest.  One has to remark that the above Hamiltonian is
invariant under a bigger symmetry than $Z_4$, namely the dihedral group
$D_4$ symmetry acting on the spin degree of freedom.  The Boltzmann
weight corresponding to (\ref{Hspin}) reads:
\begin{equation}
\exp (-H_{Z_4})=\prod_{<ij>} \left[x_0+2 x_1 \cos \frac{\pi(S_i-S_j)}{2}+x_2 \cos \pi (S_i-S_j)\right] \; ,
\end{equation}
where, by setting the normalization $x_0=1$:
\beq
x_1= \frac{\exp (A) - \exp (C)}{4}\quad  x_2=  \frac{\exp (A) -2\exp (B)+\exp (C)}{2}\; .
\eeq
The $Z_4$ spin model can be considered as a generalisation
of fourth-states Potts model, obtained by choosing $x_1=x_2$ ($B=C$). The
generalization to $x_1\neq x_2$ consists in the possibility of having
a non trivial weight between two non equal spins $S_i$ and $S_j$ which
depends on the difference between these spins, {\it i.e} $|S_i-S_j|$.

The $Z_4$ spin model (\ref{Hspin}) is a representation of the AT model
\cite{AT}.  An equivalent (and more standard) representation of the
AT model is in terms of two coupled Ising models. In this
Ising representation, on each site $i$ of a square lattice one
associates a pair of spins, denoted by $\sigma_i$ and $\tau_i$, which
takes two values, say up(+) and down(-). The Hamiltonian is defined by
\begin{equation}
H_{AT} =-\sum_{<ij>} K (  \sigma_i \sigma_j + \tau_i \tau_j )  + K_4 \sigma_i \sigma_j \tau_i \tau_j \; .
\label{HAT}
\end{equation}
In this representation the two parameters, $K$ and $K_4$, correspond
respectively to the usual Ising spin interaction and to the 4-spins
coupling between two Ising models. One can pass from the
representation (\ref{Hspin}) to the (\ref{HAT}) via the
correspondence:
\bea
S_i=1 \rightarrow \sigma_i = + , \tau_i = + \; \; ; \; \;  S_i=2 \rightarrow \sigma_i = + , \tau_i = - \nn \\ \nn
S_i=3 \rightarrow \sigma_i = - , \tau_i = - \; \; ; \; \;  S_i=4 \rightarrow \sigma_i = - , \tau_i = + \; ;  
\eea
from which one can derive  the following relation between $(K,K_4)\to (x_1,x_2)$:
\begin{equation}
\exp (4 K)={1+2 x_1 + x_2 \over 1 - 2 x_1 + x_2} \quad ; \quad \exp (2 K+ 2 K_4)=\frac{1+2 x_1 + x_2 }{1-x_2} \; .
\end{equation}
The AT model on the square lattice is equivalent to the
staggered six vertex model. It presents a rich phase diagram which has
been very well studied \cite{baxter,N}~: in particular the phase diagram
shows a critical line which is defined, in the Ising representation, by
the self-dual condition $\sinh 2 K = \exp (-2 K_4)$ and terminates at
$\coth 2 K_2=2$.  By imposing the self-dual condition, the
AT model can be solved by mapping it to the so called $F$
model, a special case of a solvable six vertex model \cite{N}.
 
On the critical line one can identify three particular points : i) the
fourth-states Potts model $K=K_4$ corresponding to
$x_1=x_2=1/3$), ii) the case of two
decoupled critical Ising models, corresponding to
$\sqrt{x_2}=x_1=-1+\sqrt{2}$ (where $K_4=0$)
and iii) the so called Fateev Zamolodchikov (FZ) point defined by
\beq
 x_1^{\mbox{FZ}}= {\sin({\pi \over 16}) \over \sin({3 \pi\over 16} )} \; 
; \; x_2^{\mbox{FZ}}= x_1 {\sin({5 \pi \over 16}) \over \sin({7\pi \over 16})} \; .
\label{FZ} 
\eeq
The FZ point has been shown to be completely integrable
\cite{FZ}. The continuum limit of the AT model on the critical line
is described by a free Gaussian field with action $\mathcal
S=\frac{1}{4\pi}\int dz d\bar{z} \partial \phi \bar{\partial}\phi$
where the scalar field $\phi$ is compactified on a orbifold of radius
$r^{orb}$, i.e.  $\phi=\phi+2\pi r^{orb}$ and $\phi=-\phi$.  The
critical Potts, (Ising)$^2$ and FZ point corresponds respectively to
$r^{orb}=2,\sqrt{2}$ and $r^{orb}=\sqrt{3}$.

The Gaussian theory with $r^{orb}=\sqrt{3}$ describes the FZ point and 
coincides with the $Z_4$ parafermionic field theory \cite{Ginsparg}.
In this paper we will mainly focus on the geometric critical
properties of the FZ point.

\section{Classification of ${\cal Z}_4$ boundary states and their spin representations}
In this section we discuss the problem of the classification of the
boundary states for the ${\cal Z}_4$ parafermionic theory.  In
particular we are interested in the representations of such states in
terms of spin configurations.  The reason is that we want to study
interfaces which are generated by imposing special boundary spin
configurations.  A classification of conformal boundary states in
terms of spin configuration can then be used to identify interfaces
whose measure is conformally invariant.

The relation between certain
spin configurations on the boundary and the conformal boundary states
is known in the case of the Ising and three-states Potts model \cite{Cardy_bcft2,bauer,AOS}. 
This is in general not true for the ${\cal Z}_4$ theory (and general
${\cal Z}_N$ theory, $N\geq 4$), where only the spin representations of a
small subset of boundary states has been explicitly discussed \cite{lukyanov}.  On the other hand, we mention that a complete
characterization of parafermionic boundary states in terms of A-D-E
lattice models degree of freedom has been accomplished in \cite{MP}.

\subsection{Rational CFTs: classification of conformal boundary conditions. }
In the following we briefly review the algebraic formulation of
boundary states for RCFTs \cite{Cardy_bcft2,ishibashi}.  A RCFT is
characterized by a certain chiral algebra and the corresponding
Hilbert space contains a finite number of the chiral algebra irrep.
$|j> $ which closes under operator product expansion.

One of the main results of the boundary RCFT
\cite{Cardy_bcft1,Cardy_bcft2} is the bijection between boundary
conformal states, which we indicate as $|\overline{j}>$, and the
irrep.  $|j>$.  In particular, the states $|\overline{i}>$ can be
expressed as:
 \begin{equation}
| \overline{i} > =\sum_{j}\frac{\mathcal{S}_{i j}}{\sqrt{\mathcal{S}_{0 j}}}  |j> 
\label{cardy_general}
\end{equation}
where the matrix $\mathcal{S}_{i j}$ determines the modular
transformation properties of the partition function defined on a
cylinder. For a RCFT the $\mathcal{S}$ matrix is in general known.
The notation $|0>$ usually indicates the (trivial) identity
representation.

We define $\mathcal{Z}_{\overline{i},\overline{j}}$ the partition
function on a cylinder with boundary conditions $|\overline{i}>$ and
$|\overline{j}>$ at the two ends. One has:
\begin{equation}
\mathcal{Z}_{\overline{i},\overline{j}}=\sum_{i} n^{i}_{\overline{i},\overline{j}} \chi_{i}(q)
\label{cil_part}
\end{equation}
where $q$ is the modular parameter, the
$n^{i}_{\overline{i},\overline{j}}$ is the number of copies of the
representation $|i>$ occurring in the spectrum and $\chi_{i}(q) $ is
the character of the representation $|i>$.  A general boundary
condition changing operator (b.c.c.)
$\psi_{\overline{0},\overline{j}}$ produces a transition from the
vacuum $|\overline{0}>$ (related to the identity representation) and
$\overline{j}$ boundary condition. The b.c.c.
$\psi_{\overline{0},\overline{j}}$ transforms in the representation
$j$. In this case the cylinder partition function reduces to a single
character as $n^{j}_{\overline{0},\overline{j}}=1$ and
\begin{equation}
\mathcal{Z}_{\overline{0},\overline{j}}=\chi_{j}(q). 
\label{single_part}
\end{equation}
The formulas (\ref{cil_part}-\ref{single_part}) are extremely useful:
given a certain lattice model, one can compute numerically the
corresponding partitions (\ref{cil_part}-\ref{single_part}) for
different boundary conditions on the cylinder. 
Then, by comparing
these results with the boundary CFT data, one can in principle
associate the CFT boundary states to specific configurations of the
degrees of freedom defining the lattice model.

We specify the above equations for the ${\cal Z}_N$ theories. 
Reminiscent of the coset construction of the parafermionic theories,
${\cal Z}_N=SU(2)_N/U(1)$, we use the notation (see \cite{maldacena})
$|l,m>$, with $l=0,1/2,1,..$ and $m$ an integer to label the primaries of
the ${\cal Z}_N$ theory. In particular the set of distinct principal
representations of the ${\cal Z}_N$ theory is given by pairs $|l,m>$ where
$l=0,1/2,1..,N/2$, $m=-2l,-2l+2..,2N-2l-2$ with $2l+m=0 \mbox{ mod }2$. The representations
$|l,m>$ and $|N/2-l,k+m>$ have to be identified.  The corresponding
boundary states $|\overline{l,m}>$ are defined by:

\begin{equation}
|\overline{l',m'}>=\sum_{l,m}\frac{\mathcal{S}^{l',m'}_{l,m}}{\mathcal{S}^{0,0}_{l,m}}|l,m>
\label{cardy_zn}
\end{equation}
where $\mathcal{S}^{l',m'}_{l,m}$ is the modular  transformation matrix \cite{maldacena} :
\begin{equation}
\mathcal{S}^{l',m'}_{l,m}=\frac{2}{\sqrt{N(N+2)}}e^{i \pi m m'/N}\sin \frac{\pi(2l+1)(2l'+1)}{N+2}\; .
\label{smatrix}
\end{equation}

\subsection{A known example: Three-states Potts model}

Before considering the $Z_4$ spin model, we would like to review the
results in \cite{Cardy_bcft2,bauer,AOS} concerning the critical $Z_3$
spin model, i.e. the critical three-states Potts model. In this case
one can identify (almost) all the boundary states in terms of spin
configurations with quite simply arguments.  Still, the $Z_3$ model is
enough rich to show some general properties of the $Z_N$ spin models
which we will try to generalize to the case $N=4$.
 
The table of principal fields of the ${\cal Z}_3$ theory with the
corresponding conformal dimension $\Delta$ is presented in
Tab.~\ref{tableZ3}.
\begin{table}
\centerline{
\begin{tabular}{c|c|c}
 Field \hspace{0.5cm} & $\Delta$ \hspace{0.5cm} & \hspace{0.2cm} $|l, m>$ \hspace{0.5cm} \\
\hline & & \\ 
$I  $\hspace{0.5cm}& 0 \hspace{0.5cm}& \hspace{0.2cm} $|0,0>$ \hspace{0.5cm} \\
& & \\
$\Psi^{1}$\hspace{0.5cm}& 2/3 \hspace{0.5cm}& \hspace{0.2cm} $|0,2>$ \hspace{0.5cm} \\
& & \\
$\Psi^{-1}$\hspace{0.5cm}& 2/3 \hspace{0.5cm}& \hspace{0.2cm} $|0,4>$ \hspace{0.5cm} \\
& & \\
$\Phi^{1} $\hspace{0.5cm}& 1/15 \hspace{0.5cm}& \hspace{0.2cm} $|1/2,1>$ \hspace{0.5cm} \\
& & \\
$\varepsilon$ \hspace{0.5cm}& 2/5 \hspace{0.5cm}& \hspace{0.2cm} $|1/2,3>$ \hspace{0.5cm} \\
& & \\
$\Phi^{-1} $\hspace{0.5cm}& 1/15 \hspace{0.5cm}& \hspace{0.2cm} $|1/2,-1 >$ \hspace{0.5cm}
\end{tabular}
}
\caption{Principal fields for the ${\cal Z}_3$ theory.
\label{tableZ3}
}
\end{table}
The fields $\Psi^{\pm 1}$ are the symmetry
currents generating the $Z_3$ chiral algebra.  It is useful to stress
that the fields listed in the above table are primaries of the
Virasoro algebra but not of the parafermionic one. Indeed one can
group the ${\cal Z}_3$ Virasoro primary fields into two families,
$\{I,\Psi^{1},\Psi^{-1}\}$ and $\{\Phi^{1},\varepsilon,\Phi^{-1}\}$
which in turn correspond to the two representation modules of $Z_3$
algebra. The fields in each module are thus connected one to the other
by symmetry transformations, or, in other words, by acting with the
modes of the $\Psi^{\pm 1}$ fields. This means that, taking into
account (\ref{smatrix}), the boundary states in
(\ref{cardy_zn}) transform under a $Z_3$ rotation as:
\begin{eqnarray}
&&|\overline{0,0}>\to |\overline{0,2}>\to |\overline{0,4}>\to
|\overline{0,0}> \nonumber \\
&&|\overline{1/2,3}>\to |\overline{1/2,1}>\to |\overline{1/2,-1}>\to |\overline{1/2,3}> \; .
\label{transf_potts}
\end{eqnarray}
This two distinct $Z_3$ ``orbifolds'' can thus be directly related to
the two already mentioned representation modules of the ${\cal  Z}_3$
parafermionic algebra.

\underline{Conformal boundary states in the spin representation}

In the three-states Potts model, where the spins can take the values
$1,2$ or $3$, the following boundary conditions have been considered
\cite{Cardy_bcft2,bauer}:
\begin{itemize}
\item {\it free}: the spins can take the values $1$, $2$ or $3$ with equal probability

\item {\it fixed}: the spins take the value $1$ or $2$ or $3$. There
  are thus three fixed boundary conditions.

\item {\it mixed}: the spins can take with equal probability the value
  $1$ or $2$ ($1+2$), $2$ or $3$ ($2+3$) , $1$ or $2$ ($1+2$).  Again
  there are three mixed boundary conditions.

\end{itemize}

The transformation (\ref{transf_potts}) greatly constraints the
possible boundary states identifications. Indeed, once one identify
the boundary state $|\overline{0,0}>$ with the fixed boundary
condition (say when the spin are fixed to the value $1$), the states
$|\overline{0,2}>$ and $|\overline{0,4}>$ have to be associated to the
other two fixed boundary conditions. Then, by observing that the free
boundary conditions are invariant under a $Z_3$ rotation, one is led
to associate the states $|\overline{1/2,1}>$, $ |\overline{1/2,3}>$
and $|\overline{1/2,-1}>$ to the other mixed boundary conditions.
Finally, taking into account that
$\mathcal{Z}_{(1|1+2)}=\mathcal{Z}_{(1|1+3)}$ (for the $Z_3$
symmetry), one can write:
\begin{eqnarray}
\mathcal{Z}_{(1|1)}&=&\chi_{I}, \quad \mathcal{Z}_{(1|2)}=\chi_{\Psi^{1}}, \quad \mathcal{Z}_{(1|3)}=\chi_{\Psi^{-1}} \nonumber \\
\mathcal{Z}_{(1|1+2)}&=&\chi_{\Phi^{1}}=\chi_{\Phi^{-1}}=\mathcal{Z}_{(1|1+3)}, \quad \mathcal{Z}_{(1|2+3)}=\chi_{ \varepsilon} \; .
\label{z3cicl}
\end{eqnarray}

To make the connection with the minimal model $M_{5}$
\cite{DiFrancesco} with central charge $4/5$ one has to take into
account relations of the kind $\chi_{I}=\chi_{(1,1)}+\chi_{(4,1)}$ or
$ \chi_{\varepsilon}=\chi_{(1,2)}+\chi_{(1,3)}$ where $\chi_{(r,s)}$
is the Virasoro character of the operator $\phi_{(r,s)}$ in the
minimal Kac table.  The boundary operator $\psi_{\varepsilon}$
transforming in the representation $\varepsilon$ generates then the
boundary conditions $(1|2+3)$ and thus the interface $SLE_{24/5}$
discussed in \cite{GC}.

One last remark: the six bulk operators which we have considered so
far do not complete the set of primaries of the ${\cal Z}_3$
parafermionic theory.  In general, the space of representation of a
${\cal Z}_N$ theory includes also fields which are associated to the
non-Abelian elements of the dihedral $D_N$ group \cite{Zamo2}.
For the ${\cal Z}_3$ theory, the two principal fields in this sector,
which we indicate as $R_{0}$ and $R_{1}$, have respectively dimensions
$1/8$ and $1/40$. The free boundary conditions have been associated
with the representation $R_{0}$ \cite{Zamo2}
\begin{equation}
\mathcal{Z}_{(1|free)}=\chi_{R_{0}},
\label{z3free}
\end{equation} 
while the representation $R_{1}$ has been associated to a {\it new}
boundary condition:
\begin{equation}
\mathcal{Z}_{(1|new)}=\chi_{R_{1}}.
\label{z3new}
\end{equation}
The physical interpretation of the new b.c.  in terms of spin
variables is actually unclear \cite{AOS}.

The conformally invariant boundary states for the three-states Potts
model are exhausted \cite{Fuchs_Schweigert} by the eight boundary
states (\ref{z3cicl}),(\ref{z3free}) and (\ref{z3new}).  Finally, one
can check these identifications by verifying that the partition
function associated to all possible combinations of such conditions
are consistent with the fusion of the principal fields. For instance, one has
\begin{equation}
\mathcal{Z}_{(1+2|1+2)}=\chi_{I}+\chi_{\varepsilon} .
\end{equation}
The above relation is consistent with the fusion $\Phi^{1}\times
\Phi^{-1}= I+ \varepsilon$, associated to the spin boundary
configuration $(1+2|1| 1+2)$ in the limit in which the two points
where the boundary conditions change approach each other.



\subsection{Spin boundary states in the ${\cal Z}_4$ spin model}

The principal primary fields of the ${\cal Z}_4$ theory are shown in Tab.\ref{tableZ4}. 
\begin{table}
\centerline{
\begin{tabular}{c|c|c}
 Field \hspace{0.5cm} & $\Delta$ \hspace{0.5cm} & \hspace{0.2cm} $|l, m>$ \hspace{0.5cm} \\
\hline & & \\ 
$I$\hspace{0.5cm}& 0 \hspace{0.5cm}& \hspace{0.2cm} $|0,0>$ \hspace{0.5cm} \\
& & \\
$\Psi^{1}$\hspace{0.5cm}& 3/4 \hspace{0.5cm}& \hspace{0.2cm} $|0,2>$ \hspace{0.5cm} \\
& & \\
$\Psi^{-1}$\hspace{0.5cm}& 3/4 \hspace{0.5cm}& \hspace{0.2cm} $|0,4>$ \hspace{0.5cm} \\
& & \\
$\Psi^{2}$\hspace{0.5cm}& 1 \hspace{0.5cm}& \hspace{0.2cm} $|0,6>$ \hspace{0.5cm} \\
& & \\
$\Phi^{1/2} $\hspace{0.5cm}& 1/16 \hspace{0.5cm}& \hspace{0.2cm} $|1/2,-1>$ \hspace{0.5cm} \\
& & \\
$\varepsilon^{'} $\hspace{0.5cm}& 9/16 \hspace{0.5cm}& \hspace{0.2cm}$|1/2,3>$\hspace{0.5cm} \\
& & \\
$\Phi^{-1/2}$\hspace{0.5cm}& 1/16 \hspace{0.5cm}& \hspace{0.2cm}$|1/2,1>$\hspace{0.5cm} \\
& & \\
$\varepsilon^{''} $\hspace{0.5cm}& 9/16 \hspace{0.5cm}& \hspace{0.2cm}$|1/2,5>$ \\
&& \\
$\Phi^{1} $\hspace{0.5cm}& 1/12 \hspace{0.5cm}& \hspace{0.2cm} $|1,2>$ \hspace{0.5cm} \\
& & \\
$ \varepsilon$ \hspace{0.5cm}& 1/3 \hspace{0.5cm}& \hspace{0.2cm} $|1,0>$ \hspace{0.5cm} \\
& & \\
\end{tabular}
}
\caption{Principal fields for the ${\cal Z}_4$ theory.
\label{tableZ4}
}
\end{table}
In this table, the fields $\Psi^{\pm 1}, \Psi^{2}$ generate,
together with the identity $I$, the $Z_4$ parafermionic
algebra. Moreover, the fields $\Psi^{2}$, together with $\varepsilon$
are the only neutral fields in the ${\cal Z}_4$ theory. In general the number
of neutral operators is equal to the dimension of the phase space of
the $Z_N$ spin model.  It is peculiar of the ${\cal  Z}_N$ theory with $N$
even, $N=2 n$, that one of these neutral fields coincides with the
current $\Psi^{n}$.

Analogously to the case of the ${\cal  Z}_3$ theory discussed above, one can
group the above fields into three representation modules,
$\{I,\Psi^{1},\Psi^{-1},\Psi^{2}\}$,
$\{\Phi^{1/2},\varepsilon^{'},\varepsilon^{''},\Phi^{-1/2}\}$ and
$\{\Phi^{1},\varepsilon\}$.  Correspondingly, the boundary states will
transform under a $Z_4$ transformation as:

\begin{eqnarray}
&&|\overline{0,0}>\to |\overline{0,2}>\to |\overline{0,4}>\to
|\overline{0,6}> \to |\overline{0,0}>\nonumber \\
&&|\overline{1/2,1}>\to |\overline{1/2,3}>\to |\overline{1/2,5}>\to |\overline{1/2,-1}> \to |\overline{1/2,1}>\nonumber \\
&&|\overline{1,0}>\to |\overline{1,2}>\to |\overline{1,0}> \; .
\label{transf_z4}
\end{eqnarray}

\underline{Conformal boundary states in the spin representation}

In the $Z_4$ model the spin can take the value $1,2,3$ or $4$. It is
natural to consider the boundary spin configurations which generalize
the $Z_3$ spin model.  Using the notations of the previous paragraph,
we consider spin boundary configurations which are the natural
extension of the one seen in the three-states Potts model:
\begin{itemize}
\item {\it free}: $1+2+3+4$.

\item {\it fixed}:$1$,  $2$, $3$ or $4$.
 
 \item {\it mixed}: Three types of mixed boundary conditions: a) $1+2$, $2+3$, $3+4$, $1+4$,  b)$1+3$, $2+4$ and c) $1+2+3$, $2+3+4$,$3+4+1$, $1+4+2$.
 
 \end{itemize}
  
 Contrary to the case of the three-states Potts model, the $Z_4$
 transformations of the boundary states, consistent with
 (\ref{transf_z4}) and (\ref{cardy_zn}) are not sufficient to identify
 all the boundary states. We have thus computed numerically the
 partitions (\ref{single_part}). The numerical implementation is done
 with a transfer matrix. In order to compute $\mathcal{Z}_{(A|B)}
 \simeq \chi_\Delta$, it is convenient to work with an infinitely long
 strip of width $L$ with boundary conditions $A$ and $B$
 \cite{Cardy_bcft2}. The corresponding dominant character
 $\chi_\Delta$ is identified by computing the subdominant corrections
 in~:
\beq
\log{(\mathcal{Z}_{(A|B)})} \simeq a_0 + {a_1 \over L} + \pi {(c/24-\Delta) \over L^2}  \; .
\eeq
Here $a_0$ is associated to the bulk free energy while $a_1$ is a boundary term corresponding to the fixed boundary conditions $A$ and $B$. 
We did the following identifications:
\begin{eqnarray}
\mathcal{Z}_{(1|1)}&=&\chi_{I}, \quad \mathcal{Z}_{(1|2)}=\chi_{\Psi^{1}}, \quad \mathcal{Z}_{(1|4)}=\chi_{\Psi^{-1}} \quad \mathcal{Z}_{(1|3)}=\chi_{\Psi^{2}}\nonumber \\
\mathcal{Z}_{(1|1+2)}=\chi_{\Phi^{1/2}}&=&\chi_{\Phi^{-1/2}}=\mathcal{Z}_{(1|1+4)}, \quad \mathcal{Z}_{(1|2+3)}=\mathcal{Z}_{(1|3+4)}=\chi_{\varepsilon^{'}}=\chi_{\varepsilon^{''}} \; .
\label{identification_z4} 
\end{eqnarray}
The mixed boundary conditions of type a) have been discussed in
\cite{lukyanov} where the infra-red behaviour of a ${\cal Z}_N$ theory
with free boundary conditions has been studied. In \cite{lukyanov},
the RG flow is characterized by a flow from the free boundary
conditions to the four stable fixed points associated to the spins
taking a fixed value on the boundary.  These results generalize the
boundary RG flow studied in \cite{AOS}. With a fine tuning of the
boundary perturbation parameters \cite{lukyanov}, the free b.c. flows
to the others four mixed boundary states of type a).

The identifications of the spin boundary states associated to
$|\overline{1,0}>$ and $|\overline{1,1}>$ remain still
ambiguous. Indeed our analysis could not discriminate between the
possible identifications:
\begin{equation}
\mathcal{Z}_{(1|1+3)}=\chi_{\Phi^{1}} \quad \mathcal{Z}_{(1|2+4)}=\chi_{\varepsilon}
\label{identification_z4_b}
\end{equation}
or
\begin{equation}
\mathcal{Z}_{(1|1+2+4)}=\chi_{\Phi^{1}} \quad \mathcal{Z}_{(1|2+3+4)}=\chi_{\varepsilon} \; .
\label{identification_z4_c}
\end{equation}
Below in the paper we will discuss the boundary conditions $(1+2|3+4)$
which, as we will see, are related to the interface studied in
\cite{CLR}.  We have verified numerically that
$\mathcal{Z}_{(1+2|3+4)}\simeq \chi_{\epsilon}$. This is consistent
with the identifications (\ref{identification_z4}) and with the
operator fusion $\Phi^{1/2} \epsilon^{'}=\epsilon+..$, determined by
the operator algebra \cite{maldacena}.  Note that the identification
of the b.c.c. operator transforming in the $\varepsilon$
representation as the one which generates the condition $(1|2+3+4)$
has been proposed in \cite{R}.  In particular this would support the
fact that the b.c.c operator transforming as $\varepsilon$ is related
to such boundary conditions \cite{CLR}.

One last remark: analogously to the $Z_3$ case, besides the fields
shown in Tab.~\ref{tableZ4}, related to the abelian $Z_4$ sector of
the theory, there are a set of fields $R$ related to the $Z_2$
reflections of the dihedral $Z_4$ group. We have verified that
$\mathcal{Z}_{1,free}=\chi_{R_0}$, where $R_{0}$ is the (twist) field
with dimension $3/24$ considered in \cite{R}.

\section{Boundary Interfaces.}

We will first present the results obtained by measuring interfaces
connecting two points of a lattice in analogy with the chordal SLE
interfaces defined in the critical $O(n)$ model. As already discussed
in section 2, the $Z_4$ spin model generalize the fourth-states Potts
model. In the Potts model, one can in general consider either
geometrical interfaces bounding spin clusters \cite{GC}, {\it i.e.}
group of spins with the same value, or Fortuin Kastelyn (FK) clusters
\cite{GR,ASZ}.  In \cite{PSS}, we explained that only geometrical
interfaces can be defined for the $Z_N$ spin models since the FK
clusters do not percolate at the critical point.

We generated these geometric interfaces by simulating finite square
lattices of size $L \times L$ with certain boundary conditions $(A_1 +
A_2 \cdots|B_1 + B_2 \cdots)$. This notation means that we set one
half of the boundary spins to take the values $A_1+A_2+..$ and the
other half the values $B_1+B_2+..$.  Moreover, we impose that the
change from one condition to the other is on the middle of the two
opposite borders of the square lattice. Then, for each spin
configuration, there is an interface defined as the line on the dual
lattice separating the $A_i$ spins connected to one boundary from the
$B_i$ spins connected to the other boundary.  

The result for the
interface associated to the condition $(1|2+3+4)$ was already
presented in \cite{PS} where we found a value of the fractal dimension $d_1$
compatible with the value $ 1 + 10/24$ predicted in
\cite{R}. Note that, as we mentioned in the previous section, 
the scenario proposed in \cite{R} is based on
the hypothesis that the b.c.c operator associated to $(1|2+3+4)$
transforms as in the representation $|1,0>$ of dimension $1/3$.
Naturally, this would be consistent with the identification
(\ref{identification_z4_c}). In order to have a more general picture,
we have systematically analyzed boundary interfaces associated to
differed boundary conditions, mainly inspired by the classification
discussed in the previous section.

Before presenting the main results, we briefly explain how the fractal
dimension of these interfaces has been obtained. In order to generate
the configurations, we choose the Ising representation (\ref{HAT}) for
which it is possible to use a cluster algorithm \cite{WD} much more
efficient than standard Monte Carlo \cite{PSS}.  For each type of
boundary conditions, measurements were performed on systems up to the
linear size $1280$. For each sizes we first determined the
autocorrelation time $\tau(L)$ and we average over $N(L) \times
\tau(L)$ with $N(L) = 1\ 000\ 000$ for $L < 160$, $N(160)= 500\ 000$,
$N(320)=250\ 000$, $N(640)=100\ 000$ and $N(1280)=50\ 000$.  A typical
value for the autocorrelation time is $\tau(L=640) \simeq 10 000$ for
the boundary condition $(1|2)$ and of the same order for other
boundary conditions and the same linear size.  The fractal dimensions
are obtained in the following way. For each linear size $L$, we
determine the average length of the interface $l(L)$ which should
scale as
\beq
l(L) \simeq L^{d_f} \; .
\eeq
Then we obtain the effective dimension as 
\beq
d_f\left({L_1 +L_2 \over 2}\right) = {\log(l(L_1)/l(L_2)) \over \log({L_1 / L_2}) } \; ,
\eeq
where $L_1$ and $L_2$ are two different linear sizes of the square lattice.

\begin{figure}[h]
\epsfxsize=240pt\epsfysize=180pt{\epsffile{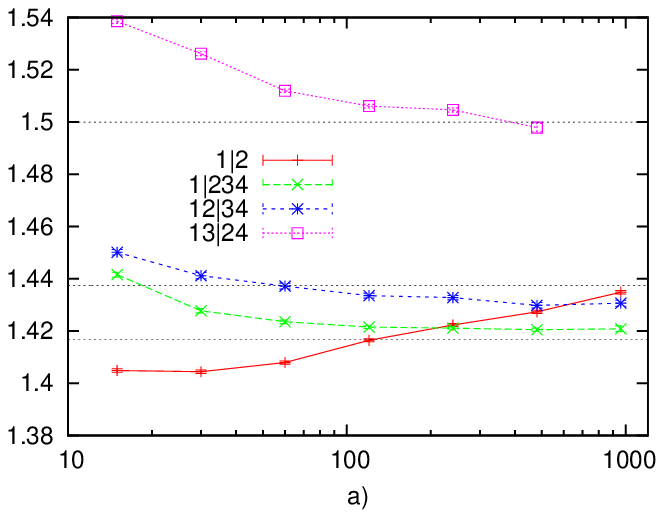}}
\epsfxsize=240pt\epsfysize=180pt{\epsffile{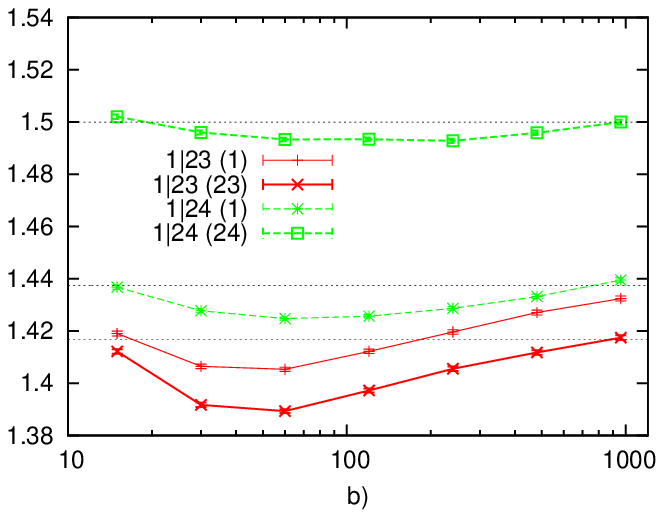}}
\caption{Effective exponents obtained for different interfaces. The three dashed lines correspond to 
$1 + 10/24,  1 + 7/16, 3/2$ as discussed in text.
\label{Plot1}
}
\end{figure}

In Fig.~\ref{Plot1}a) we present the effective fractal dimensions so
computed for the four type of interfaces associated to the
$(1|2+3+4)$, $(1+3|2+4)$ and $(1+2|3+4)$ and $(1|2)$ boundary
conditions.
The conditions $(1+2|3+4)$ or $(1+3|2+4)$ are particularly interesting
as they are, besides the condition $(1|234)$, the only other ways for defining a
single interface for this model.  As one can see from
Fig.~\ref{Plot1}a), we measured a value for the fractal dimension of the
interface $(1+2|3+4)$ which is slightly bigger than the one measured
for the $(1|234)$ interface. Even with the large amount of statistics
that we have accumulated for this point, it is not possible to decide
if the asymptotic value will converge to the same value as for the
condition $(1|2+3+4)$ or not. Note that the condition $(1+2|3+4)$ was
also considered in the recent work who obtained similar
results, namely $d_f=1.4226(13)$ \cite{CLR,mohammad}. 

On the contrary, for the condition $(1+3|2+4)$ one clearly sees that
the value is completely different and seems to converge to the value
$d_2 \simeq 3/2$. Few comments are in order.  For this condition we do not
have data for larger sizes.  This is due to the fact that the cluster
algorithm is much less efficient for studying this interface.  To be
more precise, the cluster algorithm is not able to take into account
the $(1+3|2+4)$ boundary conditions since the algorithm works for each
copy of the Ising model separately \cite{WD}. Thus one can only change
a spin $S=1 (\sigma= + ; \tau=+ ) $ in either $S=2 (\sigma= + ; \tau
=-)$ by updating the copy $\sigma$ or in $S=4 (\sigma = - ; \tau = +
)$ by updating the copy $\tau$. The direct change $S=1 \rightarrow
S=3$ can not be done within this algorithm. Since on one border the
spins $S=1$ or $3$, then these spins will be frozen, the same being
true for the other part of the border where we impose $S=2$ or
$4$. Thus for this type of boundary conditions, we consider a more
complicated algorithm in which we alternate cluster update with
standard Monte Carlo updates on the border.  Another intriguing
observation concerns the obtained value $d_2 \simeq 3/2$.  In fact, for
this choice of boundary conditions, a fractal dimension compatible
with $d_2 \simeq 3/2$ is obtained all along the AT line,
\cite{PS2}. Moreover, in the next section, we show that we find bulk
interfaces with fractal dimension $d_2$.  The fact that we obtain the
value $d_2$ appear to us natural, even if we do not have any strong
argument.  Indeed it is well known that the value $d_2$ is the fractal
dimension of certain interfaces related to the level lines of the
Gaussian free field compactified on a circle
\cite{schramm,Carduslekp,christian}. As we have seen in the section 2, the
$Z_4$ spin model on the critical line is also described by a $c=1$ CFT
which is the free Gaussian field compactified on the $Z_2$ orbifold,
$\sqrt{2}\leq r^{orb}\leq 2$.  It is thus natural to ask whether these
interfaces $(1+3|2+4)$ can be somehow related to an $SLE$ process with
$\kappa=4$.

We also report here  other cases of boundary conditions. They are
$(1|2), (1|3), (1|2+3)$ and $(1|2+4)$.  The cases $(1|2)$ and $(1|3)$
generate  two interfaces: an interface  separates the spins connected to the
boundary "1" on one side and spins $2,3,4$ on the other side while the 
second interface  separates spins $2,3,4$ with spins connected to the
boundary $2$ or $3$.  By symmetry arguments, one can see that each of
these two interfaces is equivalent \footnote{In \cite{GC} similar
  boundary conditions were considered for the three-states Potts model
  and called "fixed". In this study, the authors separate the
  interfaces in a "composite" part and a "split" part and claimed that
  the fractal dimension associated to the "split" part ($\simeq
  1.589$) was much larger than the one expected for spin cluster
  boundaries and obtained for "fluctuating" boundary conditions
  ($1+10/24 \simeq 1.41667 $). In fact the difference in these fractal
  dimensions is due to the existence of strong finite sizes
  corrections for the "split" part with "fixed" boundary conditions as
  we have checked for the three-states Potts model and for the $Z_4$ spin
  model \cite{PS2}.}. In  Fig.~\ref{Plot1}a), we show
the effective fractal dimension for the case $(1|2)$ and we average
other the two interfaces. The result for the case $(1|3)$ is identical
except finite size corrections for the smaller sizes, thus we do not
show the fractal dimension for this case.  The value obtained for the
largest sizes is $d_f \simeq 1.435 $ which is clearly different from
$d_1$.  We observe that this value, that we will call $d_3$, is very close to the value $1
+ 7/16$.  In \cite{RC,R2C,IC}, it was shown that the interfaces
related to an holomorphic operator with spin $s$ are described by
$SLE_{\kappa^{(s)}}$ with $\kappa^{(s)}=8/(1+s)$.  Here we just notice
that the value $1
+ 7/16$ corresponds to $1+\tilde{\kappa}/8$ with
$\tilde{\kappa^{(s)}}=16/\kappa^{(s)}$ and $s=3/4$ which is the
dimension (spin) of the holomorphic current $\Psi^{\pm 1}$.  As seen
in section 3, the parafermionic current $\Psi^{\pm 1}$ is associated
to the $(1|2)$ boundary conditions. 
This value is also compatible with the value
reported for bulk fractal dimension in \cite{PSS}. We will come back
on this point in the next section.  The last two cases are a little
bit more complicated. For $(1|2+3)$ we have again two interfaces but
now these two interfaces do not need to be equivalent. The first
interface separates $1$ from $234$ while the second interface
separates $23$ from $14$. In Fig.~\ref{Plot1}b), we
show the fractal dimension for each interface. It is clear that they
are not equal. For both interfaces, we observe strong finite
size effects, but the difference between the two fractal dimensions
remains near constant as we increase the linear size $L$. And it seems
that the fractal dimension associated with interface bounding the
domain connected to the boundary with $S_i=1$ converges towards $d_3$
while the fractal dimension of the other interface converges towards
$d_1$.  For the last case $(1|2+4)$ we observe again two fractal
dimensions.  The one associated to the interface bounding the domain
connected to the boundary with $S_i=1$ converges towards $d_3$ again,
while the second fractal dimension converges towards $d_2$.
 
To summarize the numerical findings, we obtain fractal dimensions
which can be grouped in three parts. A first set of fractal dimensions
converge to a value close to $d_1 \simeq 1 + 10/24$ and are associated to
either $(1|2+3+4)$ and one of the fractal dimension of $(1|2+3)$.  A
second set of fractal dimensions converge towards $d_2 \simeq 3/2$ and are
associated to $(1+3|2+4)$ and one of the fractal dimension of
$(1|2+4)$.  The last set of fractal dimensions converge to a value
close to $d_3 \simeq 1 + 7/16$ and are associated to the fractal dimension
of $(1|2)$ or the second fractal dimension for $(1|2+3)$ and
$(1|2+4)$.  

Finally the fractal dimension for $(1+2|3+4)$ takes a value between
$d_1$ and $d_3$, it is difficult to conclude definitely for this case.

\section{Bulk interfaces}

In \cite{PSS} we determined the fractal dimension of interfaces around
finite clusters in the bulk. This measurement was done by considering
any type of clusters, {\it i.e.} a geometrical cluster of spins of any
fixed value surrounded by spins with a different value. But since we
observed in the previous section that the fractal dimension can
depend on the type of interface and on particular in the number of
allowed values on each side, we will check now if a similar property also
occur for finite size clusters. The goal is to check if the fractal
dimension can depend on the number of allowed values of the geometrical
clusters and how.  In our simulations, we considered a square lattice
with periodic boundary conditions. For a given geometrical cluster we
computed the average area $A$ defined as the number of spins inside a
contour of length $l$. These two quantities are related to the fractal
dimension in the following way :
\beq
A(l)=l^{\delta}
\label{al}
\eeq
with $\delta = 2/d_f$.
To obtain this last relation, one note that $A(l)=R^2$ with 
$R$ the radius of gyration which is related to the length of the contour 
by $l=R^{d_f}$. 

We can then perform a direct measurement of the distribution of the
clusters in function of their length recording for each length the
average area. In order to compare the different type of clusters, we
first use the global $Z_4$ symmetry for each configuration. We rotate the spins such that the 
majority of them take the value "1". Next we compute all the
type of spin clusters. The simplest case is the clusters ``1'' which
corresponds to spins of value ``1'' surrounded by spins taking another
value. We can then define the associated distribution $D_1(A,l)$ which
counts the number of such clusters with area $A$ and length of the
surrounding interface $l$. One can define three one-spin
distributions, $D_1, D_2, D_3$ (by symmetry, the distribution $D_4$ is
equivalent to $D_2$).  Next we have to consider two-spins
distributions corresponding to clusters of spins taking two fixed
values surrounded by spins taking the remaining values. We have to consider $D_{12},
D_{13}, D_{23}, D_{24}$ and similarly for the three-spins, we have to consider
$D_{123}, D_{124}, D_{234}$.  For each of these distributions, one can
then extract a fractal dimension by using eq.(\ref{al}). 
Finally in order to get a precise measurement of $\delta$,
we compute the integrated quantity corresponding to
\beq
X(l_{max})= \int^{l_{max}} A(l) \simeq l_{max}^{1+\delta} \; .
\label{int}
\eeq
The values of $d_f = 1/\delta $ is 
obtained from the ratio $X(l_{max})/X(l_{max}/2)$ for increasing
$l_{max}$ for the different types of clusters.

A first result is that the fractal dimension is the same for any type
of cluster with equal sign even if we have introduce an explicit
breaking of symmetry by imposing the majority rule. We will only
consider the case of $D_2$ in the following. The same result is also
obtained for the clusters with spins restricted to three values for
which we will only consider the case of $D_{234}$. For the clusters
with spins restricted to two values, we observe two behaviors, one for
$D_{12}$ with the same result as for $D_{23}$ and a second one for
$D_{13}$ with the same result as for $D_{24}$.

\begin{figure}[h]
\epsfxsize=400pt\epsfysize=300pt{\epsffile{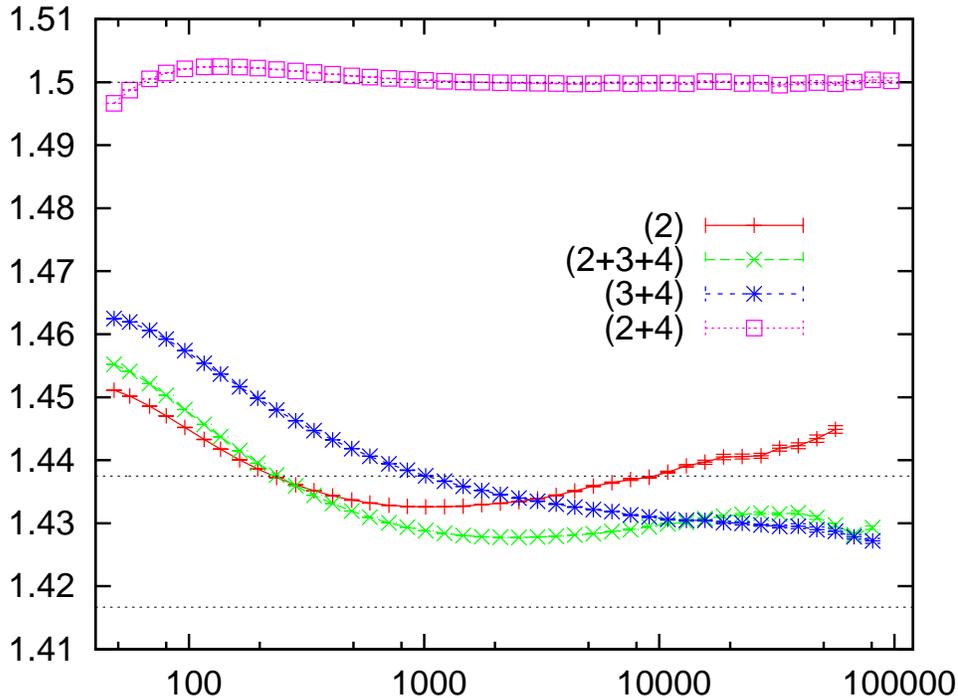}}
\caption{Effective exponents in the bulk obtained from eq.(\ref{int}) for $Z_4$.
\label{Plot2}
}
\end{figure}
In Fig.~\ref{Plot2}, we present the obtained fractal dimensions versus
$l_{max}$. This figure contains many similarities with
Fig.~\ref{Plot1}a). Again we obtain a fractal dimension $~1.5$
associated with clusters containing spins of value $1,3$ or $2,4$.
This fractal dimension is in very good agreement from the one obtained
for the interface $(1+3|2+4)$. The second fractal dimension is
associated to clusters with one value ($(2)$) and seems to be in
correspondence with the interface $(1|2)$. It is also natural to
associate the fractal dimension for clusters with three values
($(2+3+4)$) with the one from the interface $(1|2+3+4)$. The last one
$(1+2)$ or $(3+4)$ is then associated to the fractal dimension of the
interface $(1+2|3+4)$. The interesting observation is that, in the
bulk, the fractal dimensions for the clusters with three values seem
to coincide with the one for ($1+2)$ in the large size limit. This
would then confirm that the two fractal dimensions for the interfaces
$(1|2+3+4)$ and $(1+2|3+4)$ would coincide in the large size limit.

An important observation is that the fractal dimension $d_2 \simeq
3/2$ is associated to objects arising naturally in the high
temperature expansion of the AT Model \cite{Nienhuis_CG,N,Saleur_AT}. Indeed, by
replacing in (\ref{HAT}) $(\sigma_i,\tau_i)$ by $(\sigma_i,
t_i=\sigma_i \tau_i)$ then clusters of either $t_i = +1 $ or $t_i =
-1$ are natural objects to consider. Indeed, as shown in
\cite{Nienhuis_CG,N,Saleur_AT}, after performing an Ising high-temperature
expansion, the partition function defined with the Hamiltonian
(\ref{HAT}) represses as
\beq
Z_{H_{AT}} \simeq \sum_{\hbox{graphs}} (\tanh{2 K})^{l+d} \; , 
\eeq
with the sum running on two types of (non intersecting) graphs :
polygons ${\cal L}$ on the lattice coming from the expansion of
$\sigma_i$ and polygons ${\cal D}$ on the dual lattice coming from the
low-temperature expansion of $t_i$ and the total numbers of bonds on
each lattice is $l$ and $d$ respectively. The boundary of the clusters
forming ${\cal D}$ corresponds to the bulk interface with the fractal
dimension $d_2$.

\section{Summary and conclusions}
In this paper we have considered the $Z_4$ spin lattice model at the
FZ critical point.  The FZ point is described in the continuum limit
by the parafermionic ${\cal Z}_4$ CFT which is a rational CFT with
extended $Z_4$ symmetry.  Contrary to the case of the CFTs based on
Virasoro algebra, the behavior of geometric objects for general
extended RCFTs is far to be understood. In order to provide new
insights into this problem, we have considered different types of
boundary and bulk spin cluster interfaces which can be naturally
defined in the $Z_4$ spin lattice model. In particular, we have
computed their fractal dimensions at the FZ point.  The basic idea
behind the analysis presented here and in previous works
\cite{PS,PSS,R} is that, analogously to the critical $O(n)$ models
described by Virasoro CFT's, the geometrical properties of a $Z_4$ spin
lattice model at the FZ point should be related to the classification
and properties of the primary operators of the corresponding extended
algebra, in this case the parafermionic algebra.

First we considered the lattice model on a bounded domain. We studied
interfaces which origin and terminate at boundary points and which are
generated by imposing certain boundary spin configurations. We
examined all kind of boundary spin configurations whose definition is
naturally suggested by the $Z_4$ spin symmetry of the model. Despite
the great number of interfaces we considered, the numerical results
indicate that there are only three values of fractal dimension,
$d_1,d_2$ and $d_3$, characterizing the critical behavior of these
boundary interfaces.  We showed that this result can be somehow
understood on the basis on the classification of conformal invariant
boundary states. For this purpose, we have discussed the
identification of conformally boundary states, as predicted by the
boundary RCFT, in terms of boundary spin configurations.  We provided
evidences that the existence of this three different values can be
related to the different representation modules of the parafermionic
algebra.  For instance, the value $d_1 \simeq 1+10/24$, already measured in
\cite{PS,PSS}, is associated to the b.c.c. operator of dimension $1/3$
and compatible with the value predicted in \cite{R}.  Other
interfaces, like the one studied in \cite{CLR}, which seem to have the
same fractal dimension $d_1$, are associated to conformally boundary
conditions which in turn are connected by fusion to this operator.
This was in general checked on the basis of our identification of
conformal spin boundary states and fusion rules of the correspondent
${\cal Z}_4$ primaries.  For the values $d_2$ and $d_3$ there are no
theoretical arguments to derive them. However we stressed that i) 
the value $d_2$ obtained is very close to the fractal dimension $3/2$ of $SLE$
interfaces defined as level lines of free Gaussian fields and related
to a b.c.c. operator of dimension 1/4
\cite{schramm,Carduslekp,christian}. This is particularly interesting
as the ${\cal Z}_4$ is also a $c=1$ theory which can be described by a
free Gaussian field compactified on a orbifold; ii)the value $d_3\sim
1+(1+s)/4$ where $s=3/4$ is the dimension of the ${\cal Z}_4$
holomorphic current. It is natural to ask whether there is some
connection to the $SLE_{8/(1+s)}$ interfaces which are related to an
holomorphic operator with spin $s$ \cite{RC,R2C,IC}.

We have further investigated different interfaces which has been
opportunely defined in the bulk.  The results obtained for the bulk
interfaces confirmed the scenario emerging from the study of boundary
interfaces. Indeed, we could observe there are indeed three different
values of fractal dimension which are compatible with the values
$d_1,d_2$ and $d_3$. Interestingly the interfaces having the fractal
dimension $d_2 $ correspond to the spin cluster boundaries which
appear in the high-temperature expansion of the AT model
\cite{Nienhuis_CG,N,Saleur_AT}.

We conclude by emphasising that our results on critical interfaces at
the parafermionic point of the AT are expected to unveil general
aspects of geometrical objects of critical AT model, and thus of $c=1$
critical theories \cite{PS2}.  Therefore, as a theoretical
understanding is still lacking, we believe that the results presented
here can represent good motivation for studying critical interfaces in
free Gaussian field on an orbifold.



%
\newpage

\end{document}